\documentstyle{mn}
\input {psfig}
\def \kms {{\rm kms}^{-1}}

\def \oiii {[O{\small~III}]}
\def \oii {[O{\small~II}]}
\def \sii {[S{\small~II}]}
\def \nii {[N{\small~II}]}

\def \neiii {[Ne{\small~III}]}
\def \mgii {Mg{\small~II}}

\title[A small area faint KX redshift survey.]
      {A small area faint KX redshift survey for QSOs in the ESO
      Imaging Survey {\it Chandra} Deep Field South.}
\author[S.M. Croom et al.]
       {Scott M.~Croom$^{1,2}$\thanks{scroom@aaoepp.aao.gov.au}, S. J. Warren$^1$,
      K. Glazebrook$^{2,3}$\\
$^1$Astrophysics Group, Imperial College of Science, Technology and
      Medicine, Blackett Laboratory, Prince Consort Road, London, SW7 2BW\\
$^2$ The Anglo-Australian Observatory, PO Box 296, Epping, NSW 2121, Australia.\\
$^3$ Department of Physics \& Astronomy, Johns Hopkins University, 3400
      North Charles Street, Baltimore, MD 21218-2686, USA.\\}

\begin{document}


\maketitle
\begin{abstract}

In this paper we present preliminary spectroscopic results from a
small area faint K-excess (KX) survey, and compare KX selection
against UVX selection. The aim of the KX method is to produce complete
samples of QSOs flux-limited in the K band, in order to minimize any
selection bias in samples of QSOs from the effects of reddening and 
extinction. Using the photometric
catalogue of the ESO Imaging Survey {\it Chandra} Deep Field South (48
arcmin$^2$) we have identified compact objects with $J-K$ colours
redder than the stellar sequence, that are brighter than $K=19.5$. We
have obtained spectra of 33 candidates, using the LDSS++ spectrograph
on the AAT. Amongst the 11 bluer candidates, with $V-J<3$, three are
confirmed as QSOs.  Identification of the 22 redder candidates with
$V-J\ge3$ is substantially incomplete, but so far no reddened QSOs
have been found.  Near-infrared spectroscopy will be more effective in
identifying some of these targets.  Only two UVX ($U-B<-0.2$) sources
brighter than  $K=19.5$ are found which are not also KX selected.
These are both identified as galactic stars.  Thus KX selection
appears to select all UVX QSOs.  The surface density of QSOs in the
blue subsample ($V-J<3$) at $K\leq19.5$ is $325^{+316}_{-177}$
deg$^{-2}$.  Because identification of the red subsample ($V-J\ge3$)
is substantially incomplete, the $2\sigma$ upper limit on the density
of reddened QSO is large, $<1150$ deg$^{-2}$.  As anticipated, at
these faint magnitudes the KX sample includes several compact
galaxies. Of the 14 with measured redshifts, there are roughly equal
numbers of early and late type objects.  Nearly all the early type
galaxies are found in a single structure at $z=0.66$.

\end{abstract}
 \begin{keywords} surveys -- infrared: galaxies -- galaxies: active --
quasars: general
\end{keywords}

\section{Introduction}

\begin{figure}
\centering 
\centerline{\psfig{file=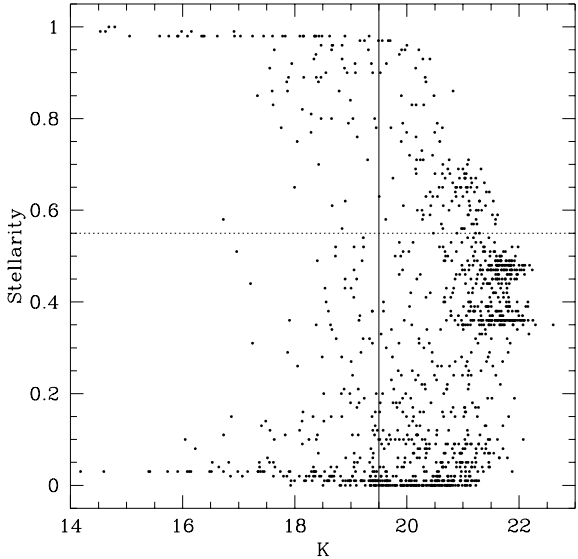,width=8.0cm}}
\caption{The stellarity index for all $K$-band sources from the
SExtractor catalogue of Rengelink et al. (1998).  The flux limit of
the sample, $K=19.5$~mag is marked with the solid vertical line.  The
stellarity limit of 0.55 is shown by the dotted line.}
\label{star_gal}
\end{figure}

The most commonly used method for finding QSOs utilizes the fact the
QSOs are bluer than most stars at ultra-violet wavelengths. This
ultra-violet excess led to the UVX technique (e.g. Schmidt \& Green
1983),\nocite{sg83} which in its simplest form uses the blue $U-B$
colours of QSOs to differentiate them from stars. This and related
methods have been utilized in most of the largest QSO surveys to date
(e.g. Durham/AAT, Boyle et al. 1990; 2dF, Croom et al. 2001). Warren,
Hewett \& Foltz (2000) have recently proposed the KX technique for
producing K-band flux limited samples of QSOs: because of the
power-law nature of their spectra, not only are QSOs bluer than stars
at UV wavelengths, but also redder than stars at near-infrared
wavelengths. In considering the KX technique it is important to
distinguish between the two effects of reddening and extinction
(e.g. Binney \& Merrifield 1998). KX QSO samples could be useful in
overcoming selection biases from both effects. First, the KX method is
insensitive to reddening. That is, the reddening vector in the $V-J$
vs. $J-K$ plane does not drive QSOs towards the stellar locus.  So the
KX method could detect QSOs that have been missed in a survey based on
UV-optical colours because they are unusually red, even though they
are brighter than the flux limit.  Second,
reddened QSOs suffer greater extinction at short
wavelengths and so can fall below the optical flux limit, but could be
detected in the K band. More generally, since
red QSOs are under-represented in surveys at short
wavelengths, and similarly blue QSOs are under-represented in surveys
at long wavelengths, it is clear that surveys at two widely separated
wavelengths are necessary to characterize fully the optical-infrared
emission from QSOs.

\begin{figure}
\centering 
\centerline{\psfig{file=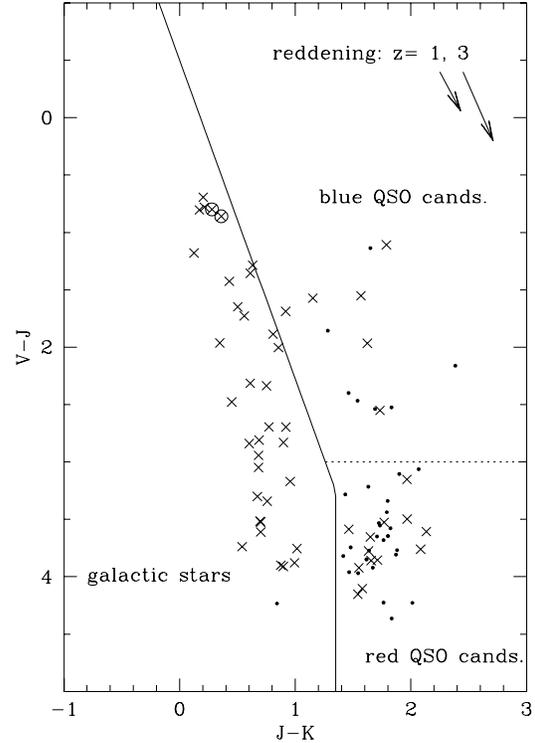,width=11.0cm}}
\caption{Colour-colour plot in $(V-J)$ vs. $(J-K)$ for $K\leq19.5$
objects with stellarity $\geq0.55$ (points).  The crosses denote
objects with stellarity $\geq0.90$.  The selection limits for our KX
sample are shown by the solid line, while the blue and red candidate
sub-samples are separated by the dotted line.  The two UVX candidates
which are not KX selected are circled.  In the top right-hand corner
we plot the reddening vectors for absorbers at $z=1$ and 3 for
$E(B-V)=0.1$ using the LMC extinction curve.}
\label{col_sel}
\end{figure}

There is currently considerable interest in quantifying the numbers of
red QSOs. Red QSOs have been discovered in radio selected samples
\cite{web95}, and highly absorbed QSOs have been invoked to explain
the spectrum of the X-ray background radiation (e.g. Madau et
al. 1994).  The unified model of AGN (e.g. Antonucci 1993) postulates a dusty
torus surrounding the central black hole engine.  At high inclination
angles, when observing through the torus, the gas associated with the
torus will absorb soft X-rays, hardening the intrinsic QSO spectrum such
that it can be reconciled with the harder X-ray background.  The dust
also preferentially extinguishes blue optical/UV flux, reddening the object
at optical wavelengths.  A small number of these ``type 2 QSOs'' have
been discovered (e.g. Almaini et al., 1995; Boyle et al., 1998) in
deep X-ray samples, showing narrow emission lines in the optical, but
broad lines in the near-infrared (IR).  One way to reduce the bias
against selecting such objects is to construct an IR selected sample
(alternatives would be to use hard X-ray or deep radio surveys).

\begin{table*}
\begin{center}
\caption{KX selected objects and spectroscopic identifications for the blue
sub-sample, ($V-J\leq3.0$).  The magnitudes are listed in the $UBVRJK$ bands and
S/G$_{\rm K}$ denotes the SExtractor stellarity parameter in the $K$ band.
Objects with an ID of ``-'' have not been observed.  Those with a ``?'' ID could
not be identified from their spectra.  An (e) or (a) after a ``gal'' ID denotes 
emission or absorption dominated spectra respectively.  The observed spectral 
features, in both absorption and emission are also listed.}
\tabcolsep=4pt
\begin{tabular}{lcccccccccccl}
\hline 
name & RA(J2000) & Dec.(J2000) & $U$ & $B$ & $V$ & $R$ & $J$ & $K$ & S/G$_{\rm K}$ & $z$ & ID &
spectral features\\
\hline
KX1  &   3 32 08.70 & --27 47 34.5 &  18.90 & 19.23 &18.64 &18.44 &17.07 &15.92 &0.98  & 0.544 &    QSO & \oii,\neiii,H$\delta$,H$\gamma$,H$\beta$,\oiii\\ 
KX2  &   3 32 09.48 & --27 48 06.9 &  22.58 & 21.67 &20.69 &20.37 &19.01 &18.09 &0.98  & 0.000 &      - &                     \\ 
KX3  &   3 32 16.53 & --27 44 49.2 &  23.13 & 23.84 &23.16 &22.73 &20.76 &19.30 &0.71  & 0.975 &    gal(e) & \oii \\ 
KX4  &   3 32 30.02 & --27 45 30.2 &  20.54 & 21.53 &21.33 &21.03 &20.22 &18.43 &0.98  & 1.221 &    QSO & \mgii,\oii\\ 
KX5  &   3 32 30.11 & --27 45 24.0 &  22.78 & 23.52 &23.02 &22.50 &20.50 &18.67 &0.61  & 0.000 &      ? &                      \\ 
KX6  &   3 32 31.51 & --27 46 23.4 &    -   & 24.17 &23.58 &22.80 &21.42 &19.04 &0.80  & 0.296 &    gal(a) & H,K,H$\delta$,G   \\ 
KX7  &   3 32 33.06 & --27 46 09.1 &  23.92 & 24.10 &22.80 &21.85 &20.33 &18.79 &0.86  & 0.000 &      ? &                      \\ 
KX8  &   3 32 33.58 & --27 46 23.8 &  22.61 & 23.05 &21.98 &21.38 &20.12 &18.84 &0.56  & 0.276 &    gal(e) & H,K,H$\delta$,G,H$\gamma$,H$\alpha$,\nii \\ 
KX9  &   3 32 38.83 & --27 47 32.5 &  21.19 & 22.11 &21.50 &20.80 &19.95 &18.38 &0.94  & 0.458 &    gal(e) & \oii,H,K,H$\delta$,G,H$\gamma$,H$\beta$,\oiii\\ 
KX10 &   3 32 39.13 & --27 46 02.1 &  20.58 & 21.55 &21.21 &20.77 &20.07 &18.42 &0.88  & 1.221 &    QSO & \mgii,\oii \\ 
KX11 &   3 32 41.80 & --27 46 19.7 &  24.01 & 23.98 &22.36 &21.40 &19.82 &18.13 &0.82  & 0.333 &    gal(e) & \oii,H,K,G,H$\gamma$,H$\beta$,H$\alpha$,\nii \\ 
KX12 &   3 32 44.24 & --27 47 33.8 &  22.46 & 23.28 &22.66 &21.80 &20.70 &19.07 &0.92  & 0.000 &      - &                     \\ 
KX13 &   3 32 45.15 & --27 47 24.2 &  22.58 & 22.99 &21.92 &20.90 &19.37 &17.64 &0.95  & 0.437 &    gal(e) & \oii,H,K,H$\delta$,H$\beta$,H$\alpha$,\nii \\ 
\hline
\label{data_table1}
\end{tabular}
\end{center}
\end{table*}

\begin{table*}
\begin{center}
\caption{KX selected objects and spectroscopic identifications for the
red sub-sample, ($V-J>3.0$).  The format is the same as that in Table
\ref{data_table1}.}
\tabcolsep=4pt
\begin{tabular}{ccccccccccccl}
\hline 
name & RA(J2000) & Dec.(J2000) & U & B & V & R & J & K & S/G$_{\rm K}$ & z & ID &
spectral features\\
\hline
KX14 &  3 32 06.48  &--27 47 28.8  & 23.53 & 23.95  & 22.89 & 22.32 & 19.67 & 18.04 & 0.75   & 0.000 &      - &                     \\ 
KX15 &  3 32 06.61  &--27 46 23.1  &   -   & 26.02  & 25.35 & 23.90 & 21.13 & 19.11 & 0.76   & 0.000 &      - &                     \\ 
KX16 &  3 32 08.70  &--27 45 02.0  & 25.45 & 25.64  & 23.63 & 22.38 & 19.71 & 18.16 & 0.98   & 0.000 &      - &                     \\ 
KX17 &  3 32 10.57  &--27 46 29.1  &   -   &   -    &   -   &   -   & 21.49 & 19.45 & 0.78   & 0.000 &      ? &                     \\ 
KX18 &  3 32 11.26  &--27 45 33.7  &   -   & 26.66  & 24.54 & 23.74 & 20.39 & 18.84 & 0.96   & 0.333 &    gal(e) &   H$\beta$,\oiii,H$\alpha$    \\ 
KX19 &  3 32 11.65  &--27 45 54.4  &   -   & 25.94  & 25.02 & 23.34 & 19.97 & 18.47 & 0.80   & 0.000 &      - &                     \\ 
KX20 &  3 32 12.24  &--27 45 30.3  & 24.77 & 24.92  & 22.60 & 21.54 & 19.32 & 17.89 & 0.89   & 0.000 &      ? &                     \\ 
KX21 &  3 32 12.50  &--27 47 29.3  &   -   & 25.61  & 24.53 & 23.31 & 20.42 & 18.84 & 0.93   & 0.000 &      ? &                     \\ 
KX22 &  3 32 13.05  &--27 46 38.2  &   -   &   -    &   -   & 24.64 & 20.85 & 19.31 & 0.95   & 0.212 &    gal(a) &  H,K,H$\delta$,G,H$\beta$,\oiii \\ 
KX23 &  3 32 14.48  &--27 46 24.8  &   -   & 25.26  & 23.94 & 22.66 & 20.36 & 18.89 & 0.98   & 0.000 &      ? &                     \\ 
KX24 &  3 32 14.82  &--27 44 33.4  & 25.73 &   -    & 25.38 & 23.84 & 21.15 & 19.39 & 0.90   & 0.000 &      - &                     \\ 
KX25 &  3 32 15.56  &--27 45 36.5  &   -   & 26.25  & 24.85 & 23.47 & 20.49 & 18.66 & 0.80   & 0.000 &      ? &                     \\ 
KX26 &  3 32 16.57  &--27 47 27.4  &   -   & 25.50  & 23.51 & 22.33 & 19.77 & 18.29 & 0.77   & 0.000 &      - &                     \\ 
KX27 &  3 32 17.23  &--27 46 49.4  &   -   & 26.89  & 24.50 & 23.13 & 20.53 & 18.99 & 0.84   & 0.000 &      ? &                     \\ 
KX28 &  3 32 17.80  &--27 47 15.1  &   -   & 25.57  & 23.23 & 22.06 & 19.41 & 18.00 & 0.65   & 0.000 &      - &                     \\ 
KX29 &  3 32 18.49  &--27 45 56.2  &   -   & 25.35  & 23.35 & 22.05 & 19.43 & 17.76 & 0.78   & 0.000 &      - &                     \\ 
KX30 &  3 32 18.90  &--27 47 34.6  &   -   & 27.11  & 24.33 & 23.13 & 20.75 & 18.93 & 0.90   & 0.000 &      ? &                     \\ 
KX31 &  3 32 19.27  &--27 46 32.4  &   -   & 25.67  & 23.32 & 22.11 & 19.47 & 17.86 & 0.88   & 0.000 &      ? &                     \\ 
KX32 &  3 32 19.90  &--27 47 21.4  &   -   & 26.47  & 23.98 & 22.73 & 20.12 & 18.47 & 0.96   & 0.000 &      - &                     \\ 
KX33 &  3 32 19.96  &--27 48 31.2  &   -   & 25.53  & 24.07 & 22.82 & 20.30 & 18.66 & 0.96   & 0.000 &      - &                     \\ 
KX34 &  3 32 20.51  &--27 47 32.5  & 25.10 & 25.24  & 23.30 & 22.37 & 20.19 & 18.29 & 0.81   & 0.000 &      - &                     \\ 
KX35 &  3 32 24.42  &--27 46 24.6  &   -   & 26.08  & 24.33 & 23.28 & 20.37 & 18.90 & 0.62   & 0.000 &      ? &                     \\ 
KX36 &  3 32 26.71  &--27 46 59.3  &   -   & 26.33  & 24.49 & 23.14 & 20.72 & 19.08 & 0.90   & 0.000 &      ? &                     \\ 
KX37 &  3 32 28.49  &--27 47 04.0  & 24.06 & 24.85  & 23.23 & 21.84 & 19.58 & 17.87 & 0.86   & 0.668 &    gal(a) & \oii,H,K,G,Mgb   \\ 
KX38 &  3 32 28.79  &--27 46 20.7  &   -   & 24.32  & 23.24 & 22.08 & 19.71 & 17.95 & 0.92   & 0.000 &      ? &                     \\ 
KX39 &  3 32 29.27  &--27 47 07.8  & 25.17 & 25.11  & 23.12 & 21.63 & 19.26 & 17.55 & 0.91   & 0.668 &    gal(a) & H,K,H$\delta$,G,H$\gamma$,H$\beta$,Mgb \\ 
KX40 &  3 32 29.99  &--27 47 57.4  &   -   & 25.63  & 24.09 & 22.96 & 21.02 & 18.96 & 0.79   & 0.000 &      - &                     \\ 
KX41 &  3 32 30.07  &--27 47 27.1  & 24.50 & 24.65  & 24.90 & 24.61 & 21.29 & 19.16 & 0.97   & 0.000 &      - &                     \\ 
KX42 &  3 32 30.89  &--27 46 22.1  & 24.02 & 24.74  & 23.96 & 23.10 & 20.30 & 18.66 & 0.95   & 0.000 &      ? & 		           \\ 
KX43 &  3 32 31.26  &--27 45 33.1  &   -   &   -    &   -   & 24.33 & 21.39 & 19.42 & 0.80   & 0.000 &      - &                     \\ 
KX44 &  3 32 32.09  &--27 44 52.1  & 24.98 & 25.48  & 24.15 & 22.85 & 20.34 & 18.47 & 0.89   & 0.000 &      - &                     \\ 
KX45 &  3 32 33.89  &--27 46 00.6  & 24.78 &   -    & 25.21 & 24.71 & 21.45 & 19.37 & 0.92   & 0.000 &      ? &                     \\ 
KX46 &  3 32 35.80  &--27 47 59.1  &   -   & 25.09  & 22.78 & 21.37 & 19.10 & 17.33 & 0.85   & 0.665 &    gal(a) & H,K,H$\delta$,G,H$\gamma$  \\ 
KX47 &  3 32 37.35  &--27 47 29.6  &   -   & 24.81  & 22.37 & 20.99 & 18.61 & 16.73 & 0.58   & 0.669 &    gal(a) & \oii,H,K,G,H$\gamma$       \\ 
KX48 &  3 32 38.30  &--27 44 55.4  &   -   & 26.08  & 23.68 & 22.28 & 20.15 & 18.43 & 0.70   & 0.666 &    gal(a) & H,K,G,H$\gamma$  \\ 
KX49 &  3 32 38.83  &--27 46 49.2  &   -   & 25.70  & 23.67 & 22.12 & 20.12 & 18.38 & 0.89   & 0.000 &      - &                     \\ 
KX50 &  3 32 38.83  &--27 44 49.3  &   -   & 25.65  & 23.81 & 22.40 & 20.31 & 18.34 & 0.93   & 0.000 &      - &                     \\ 
KX51 &  3 32 39.27  &--27 47 58.6  & 25.39 & 25.54  & 23.04 & 21.68 & 19.40 & 17.60 & 0.86   & 0.000 &      - &                     \\ 
KX52 &  3 32 40.71  &--27 47 31.2  & 23.98 & 24.76  & 23.30 & 22.05 & 19.96 & 18.16 & 0.89   & 0.667 &    gal(e) & \oii,H,K,G,H$\beta$,\oiii \\ 
KX53 &  3 32 41.45  &--27 47 17.4  &   -   & 24.93  & 22.84 & 21.53 & 19.40 & 17.61 & 0.83   & 0.000 &      ? &                     \\ 
KX54 &  3 32 47.38  &--27 46 27.3  &   -   & 26.21  & 24.61 & 23.45 & 21.46 & 19.49 & 0.97   & 0.000 &      - & \\ 
\hline
\label{data_table2}
\end{tabular}
\end{center}
\end{table*}

As well as intrinsic absorption, QSOs may also suffer extinction and
reddening by dusty objects which lie along their line of site.  For
example, a dusty damped Lyman $\alpha$ absorber
(DLA) will have the effect of making the background QSO both fainter
and redder \cite{fp93}.  A bias will then be introduced in the distribution of
DLAs detected from optically selected QSO samples.  This bias is
apparent as a lack of high H{\small~I} column (log
$N$[H{\small~I}]$>21$), high metallicity DLAs \cite{blbd98}.  A second
example is that of gravitational lenses.  QSOs behind a dusty lens are
less likely to be detected, potentially biasing low the number of
gravitational lenses found.  This will have a significant impact on
the cosmological interpretation of the number of gravitationally
lensed QSOs.

This paper reports on a test of the KX method. Candidate QSOs are
stellar sources with $J-K$ colours redder than the stellar sequence in
a $V-J$ vs. $J-K$ two-colour diagram.  Because of the small areas so far
covered by surveys in the near-infrared, to produce a QSO sample of
significant size we decided to go deep, rather than wide. The
advantage of this strategy is that we can use multi-object
spectroscopy for candidate confirmation. The main disadvantage is that
the list of candidates will be contaminated by compact galaxies, which
is not a significant problem at bright magnitudes.  This is made worse
by the fact that faint QSOs may also be resolved, so that the
position of the star/galaxy discrimination boundary must be relaxed,
in order for the sample to be complete. Section \ref{phot_sec}
describes the photometric data, and the candidate selection procedure
employed.  Section \ref{spec_sec} presents the spectroscopic
observations which were carried out with LDSS++ at the
Anglo-Australian Telescope.  We discuss our results in Section
\ref{disc_sec}.

\section{Photometric data and candidate selection}\label{phot_sec}

\begin{figure*}
\centering 
\centerline{\psfig{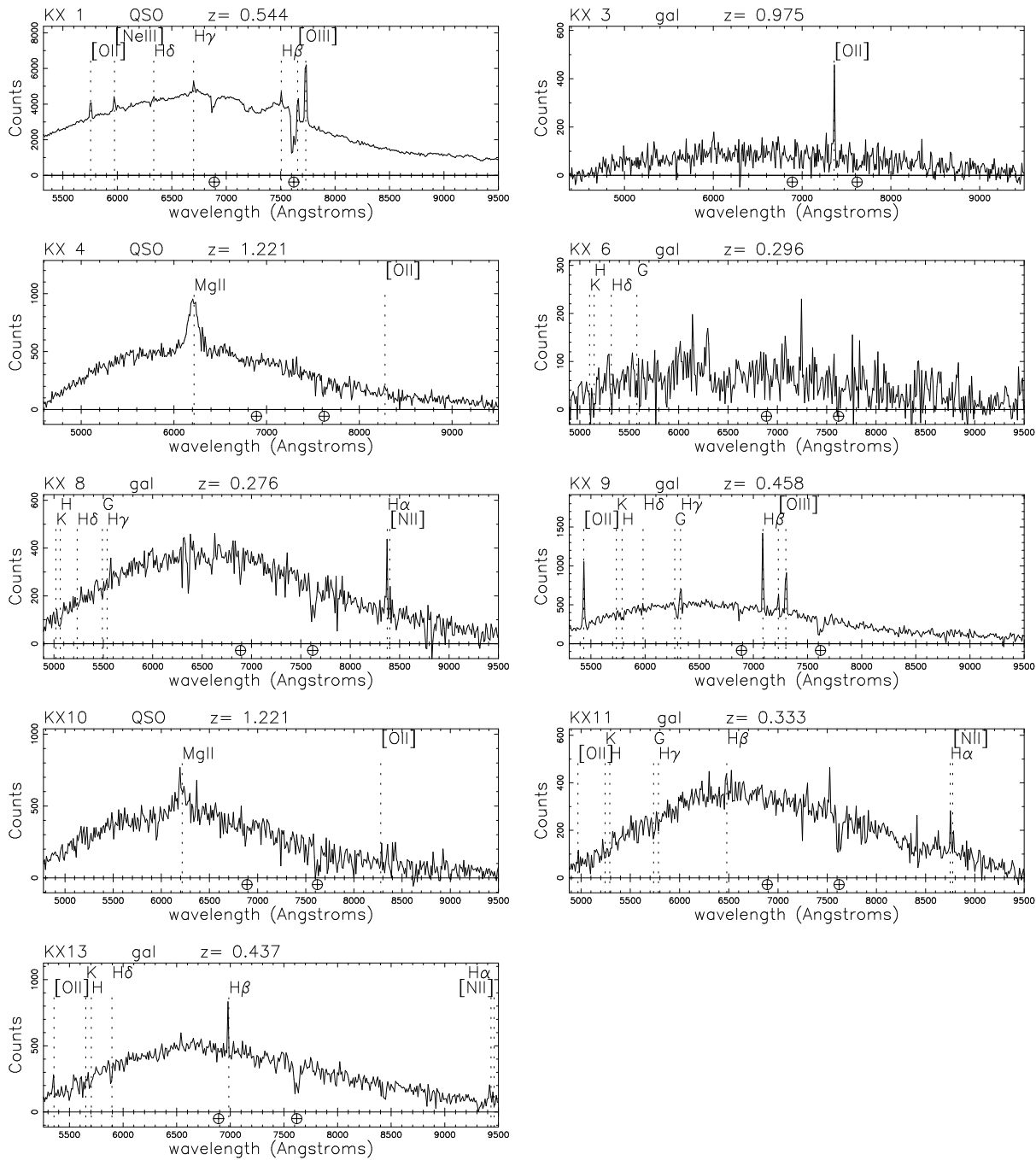}}
\caption{Spectra of the identified objects in the blue subsample.  The
wavelengths of spectral features are labelled on each spectrum.  We
have not corrected for telluric absorption bands, and indicate these
by a circled cross.}
\label{spec1}
\end{figure*}

\begin{figure*}
\centering 
\centerline{\psfig{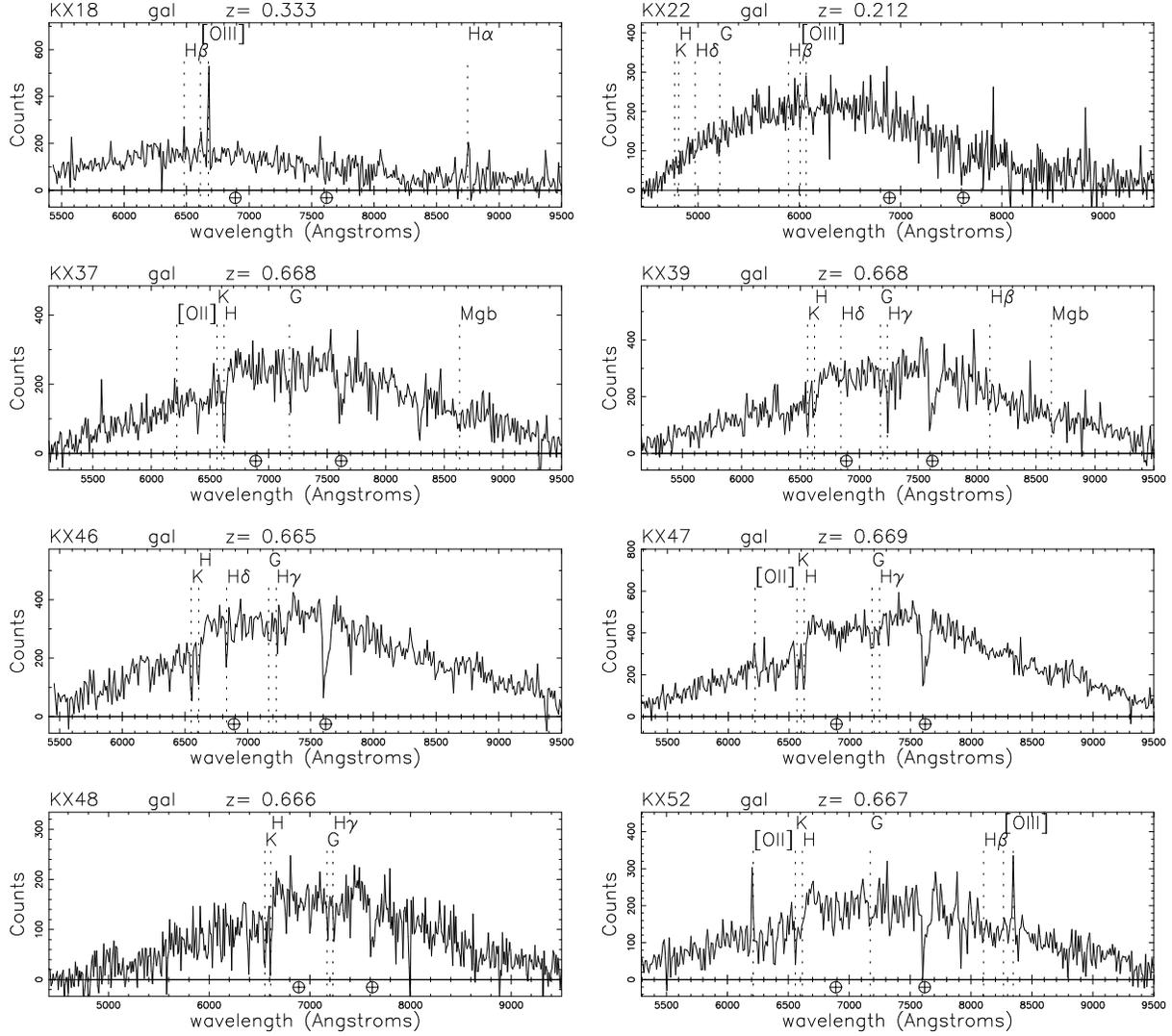}}
\caption{Spectra of the identified objects in the red subsample.}
\label{spec2}
\end{figure*}

The photometric data used was that of the ESO Imaging Survey {\it
Chandra} Deep Field South (EIS-CDFS; formally the EIS-AXAF deep field)
\cite{EIS98}. The EIS-CDFS deep field covers an area
of 48arcmin$^2$, and was observed in 7 bands $UBVRIJK$ at the New
Technology Telescope (NTT) using the SUSI2 and SOFI instruments. The I
band coverage of the field is incomplete. Rengelink et al. produced
catalogues of detections in each band using SExtractor \cite{ba96}, which generates
positions, fluxes, magnitudes and a number of other image parameters
including a star-galaxy discrimination parameter. The $2\sigma$
magnitude limits are $U\simeq25.2$, $B\simeq27.0$, $V\simeq26.5$,
$R\simeq26.3$, $I\simeq25.6$, $J\simeq23.6$ and $K\simeq21.6$. The
data are publicly available from {\tt http://www.eso.org/eis}.

Beginning with the $K$ catalogue we selected all objects brighter than
$K=19.5$, and then picked out the relevant photometry from the $V$ and
$J$ catalogues.  Unfortunately at these faint magnitudes QSOs may
appear non-stellar. Therefore the selection of the star-galaxy
discrimination boundary involves balancing the competing requirements
of completeness of the QSO sample, and contamination by galaxies.  We
have included candidates that are marginally non-stellar, applying a
cut on the K-band SExtractor stellarity parameter $\geq0.55$, where
stellarity $=1$ denotes a star and stellarity $=0$ denotes a galaxy.
Fig. \ref{star_gal} shows the stellarity index plotted as a function
of $K$ magnitude together with our selection limits.   As illustrated
in Fig. \ref{col_sel}, all but one of  the objects lying in the
stellar locus of the two-colour diagram have stellarity $\geq0.9$, so
our applied  cut is quite conservative.  We can anticipate, therefore,
that our candidate list will include several galaxies. The KX and UVX
methods have this problem in common at faint magnitudes.  The KX
colour selection was carried out in the $(V-J)$ vs. $(J-K)$ plane, as
shown in Fig. \ref{col_sel}.  Compact objects which have $(J-K)>1.35$
or $(J-K)>0.36(V-J)+0.18$ were selected for follow up spectroscopy.
This selection is almost identical to that used by Warren et
al. (2000).  We make a comparison to UVX selection by finding the
$U-B$ colours for all our compact $K<19.5$ sources.  Of those sources
with positive detections in the $U$ and $B$ bands only two are found
to be selected as UVX ($U-B<-0.2$) and not KX.  A further 18 sources
are selected as both UVX and KX.  The non-KX UVX sources
have similar colours in both the optical and IR, these are typical of
blue (e.g. A/B) stars.  They lie at the blue end of the stellar locus
in the $(V-J)$ vs. $(J-K)$ plane (the circled points in
Fig. \ref{col_sel}).  

We divided our KX candidate list into two
sub-samples, either side of $(V-J)=3$.  Table 1 lists the 13 objects
in the bluer of the two sub-samples, which lie in the same region of
the colour-colour plane as typical blue optically selected QSOs. These
candidates have average $R=21.3$. Table 2 lists the 41 objects in the
red sub-sample, containing candidate highly reddened QSOs. These
candidates have average $R=22.7$. This sub-sample is expected to be
contaminated by compact distant early-type galaxies.

\section{Spectroscopic observations}\label{spec_sec}

Spectroscopic observations were carried out with the Low Dispersion
Survey Spectrograph (LDSS++) on the nights of the 11th and 12th of
December 1999.  LDSS++ uses the MITLL3 CCD, along with a volume phase
holographic grism to give very high efficiency ($\sim25\%$ at
$7000$\AA).  Two masks were observed, each containing $\sim45$ slits
of length $6''$ and width $1.5''$.  Faint objects, where possible,
were included in both masks.  Priority was given to the blue
sub-sample, resulting in 11 of the 13 objects being allocated slits,
while a smaller fraction (22 out of 41) of the red sub-sample was
allocated slits.  We were also able to allocate a slit to one of the
two non-KX UVX candidates.  To fill in the remaining space, in positions where
there were no KX objects, we randomly selected $K\leq19$ galaxies
(stellarity $\leq0.1$). 

The nod-and-shuffle technique was used to
optimize sky subtraction \cite{gbh2001}. The principle of the method is
to quasi-simultaneously record the sky local to the object through exactly
the same light path as the object flux, so that the sky can
be subtracted. This is in contrast to the usual method which is to subtract a
smooth function fit to the sky counts at each wavelength.
The telescope was nodded back and forth every 30 seconds by $3''$,
keeping the object in the slit.  At the same time, the charge was
shuffled up and down the CCD, such that a single read out contained
two sets of spectra, one for each position of the object in the slit.
By simply subtracting one set of spectra from the other, near perfect
(i.e. Poisson limited) sky removal is obtained, as the sky flux passes
through the same part of the slit, and is incident on the same pixel
in the CCD as the object flux. Although the subtraction process
increases the sky noise by $\sqrt 2$, compared to the usual method,
the greatly reduced systematic errors allow reliable spectroscopy of
much fainter objects.

During the first night of observations we obtained $5\times2000$s
exposures on mask 1, in seeing of $\sim2''$. On the second night,
$7\times2000$s exposures were obtained on mask 1 in $1-1.5''$ seeing
and $3\times2000$s exposures were obtained on mask 2.
Copper-Helium-Neon arcs were taken at intervals throughout the night
to wavelength calibrate the spectra.

The CCD reduction, spectral extraction and calibration was carried
out within the {\small IRAF} package, using the tasks within {\small
IMRED.CCDRED} and {\small IMRED.SPECRED}.  After bias subtracting the
images and interpolating over bad pixels and columns, the images were
combined in groups of 2 or 3, averaging and using clipping algorithms
to remove cosmic ray events.  Combining in small groups, rather than
over a whole night, ensured no smearing out of the spectra due to small
shifts (typically a fraction of a pixel) caused by flexure in the
spectrograph.  Each combined frame was split in half, separating the
spectra taken at the two nod-and-shuffle positions.  One of these two
halves was then subtracted from the other to produce the sky
subtracted image containing both positive and negative spectra for
each object.  The routine {\small APEXTRACT} was then used to trace
and extract the positive and negative spectra.  A number of the
spectra have extremely low S/N and were therefore impossible to trace.
In these cases the trace used was one interpolated from surrounding
well traced spectra.  Wavelength calibration was done using the
{\small IDENTIFY} and {\small DISPCOR} routines.  Finally, multiple
spectra of the same object were combined together.

\begin{table*}
\begin{center}
\caption{The two UVX non-KX sources with $K\leq19.5$ in our survey area}
\begin{tabular}{lccccccccc}
\hline 
name & RA(J2000) & Dec.(J2000) & U & B & V & R & J & K & S/G$_{\rm K}$ \\
\hline
UVX1 & 3 32 30.58 & --27 44 59.8 & 19.75 & 20.01 & 19.56 & 19.20 & 18.76 & 18.48 & 0.98\\
UVX2 & 3 32 44.08 & --27 48 24.4 & 19.50 & 19.93 & 19.37 & 18.92 & 18.51 & 18.15 & 0.98\\
\hline												      						          
\label{data_table4}
\end{tabular}
\end{center}
\end{table*}

\begin{figure*}
\centering 
\centerline{\psfig{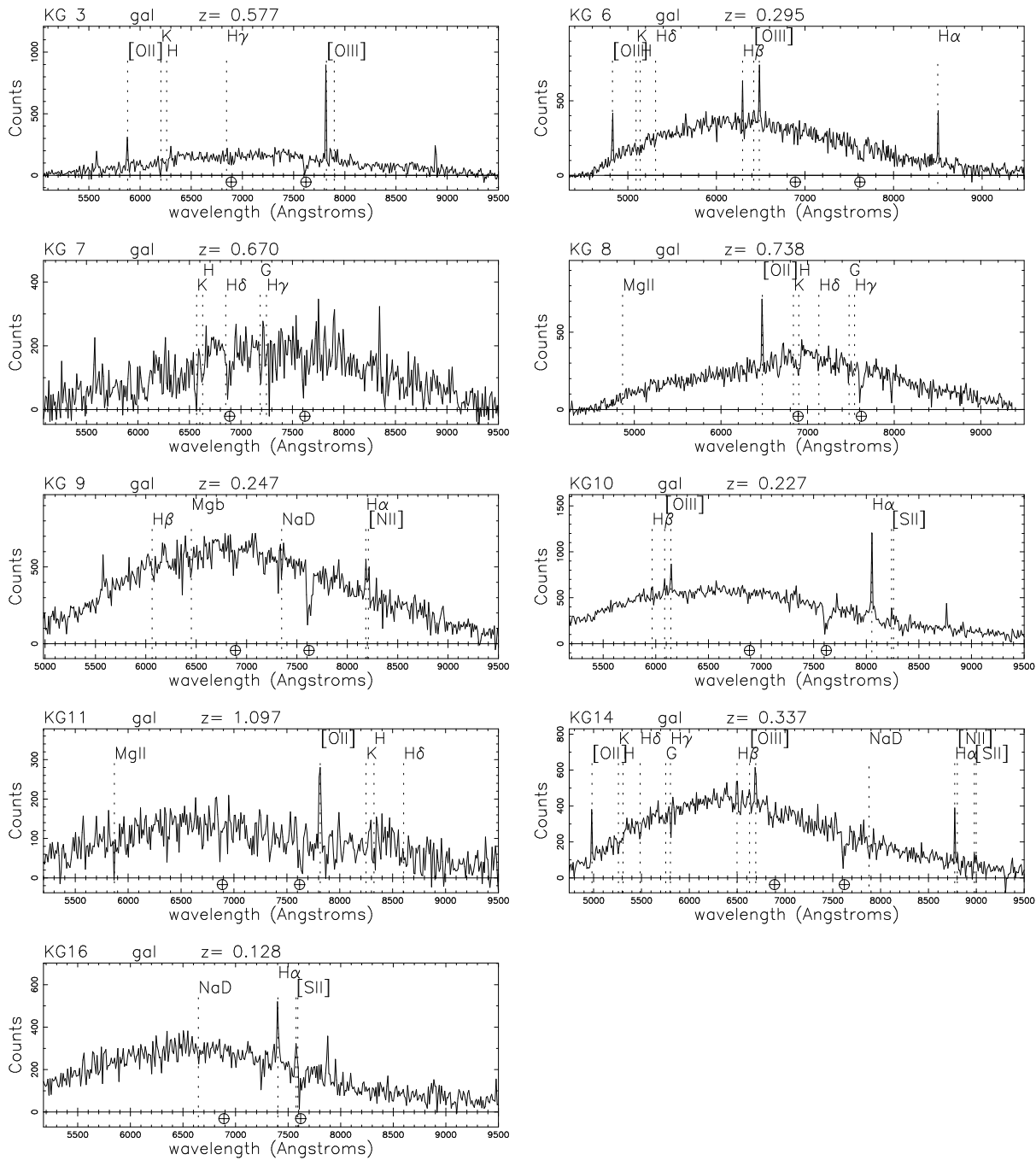}}
\caption{Spectra of the identified objects in the galaxy sample.}
\label{spec3}
\end{figure*}

Classification and redshift determination were undertaken by eye using
a modified version of the {\small REDSHIFT} program by Karl
Glazebrook.  A high fraction of the blue sub-sample could be
identified (9 of 11 observed), while the redder sub-sample had a much
lower identification rate (only 8 of 22), as expected given their much
fainter $R$-band fluxes.  The positions, magnitudes, identifications,
and redshifts are given in Tables \ref{data_table1} and
\ref{data_table2} for the blue and red sub-samples respectively, and
the spectra are shown in Figs. \ref{spec1} and \ref{spec2}.  The two
UVX selected objects are listed in Table \ref{data_table4}.  We
obtained a spectrum of UVX1 and it was found to be an A star, with
clear Balmer absorption lines.  The $K\leq19$ galaxy sample is listed
in Table \ref{data_table3} and the galaxy spectra are shown in
Fig. \ref{spec3}.

\begin{figure}
\centering 
\centerline{\psfig{file=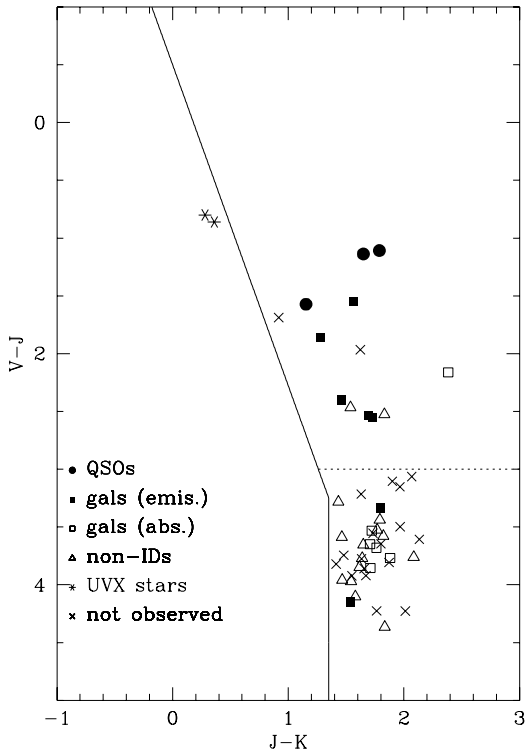,width=11.0cm}}
\caption{Colour-colour plot in $(V-J)$ vs. $(J-K)$ for our KX selected
sample.  Different symbols denote the different object IDs: QSOs (filled
circles); emission line galaxies (filled squares);
absorption line galaxies (open squares); non-IDs (open triangles);
unobserved objects (crosses).}
\label{colselid}
\end{figure}

\begin{figure}
\centering 
\centerline{\psfig{file=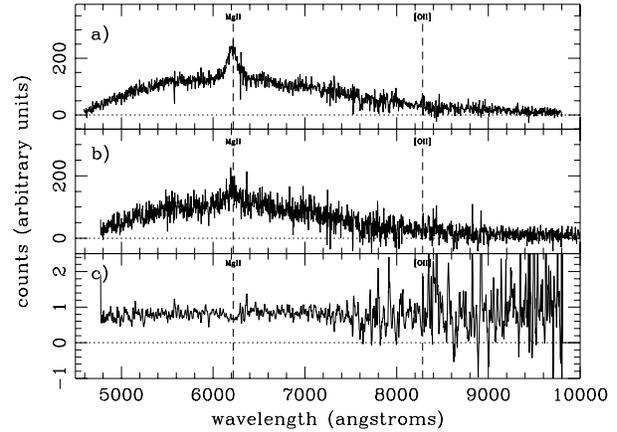,width=8.0cm}}
\caption{spectra of the QSO pair found in our KX sample KX4 (a) and
KX10 (b) with emission lines denoted by the dashed lines.  c) shows
the flux ratio KX10/KX4 boxcar smoothed with a window of 5
pixels.} 
\label{qsospec}
\end{figure}

The KX objects are plotted in the $(V-J)$ vs. $(J-K)$ plane in
Fig. \ref{colselid}.  We identify 3 QSOs; one at $z=0.544$ and a pair
of QSOs at $z=1.221$.  All 3 objects would also be selected in a UVX
sample, having colours $(U-B)=-0.33$, --0.99 and --0.97 for KX1, KX4
and KX10 respectively.  KX1 also shows narrow H$\beta$ superimposed on
a much broader underlying component.  The spectra of the $z=1.221$ QSO
pair and their flux ratio are shown in Fig. \ref{qsospec}.  This pair
has an angular separation of $125''$.  The ratio of the two spectra is
flat over the entire spectroscopic range (Fig. \ref{qsospec}c).  They
also have similar fluxes in all the optical and IR bands.  However,
the large angular separation makes this pair an unlikely candidate for
a gravitational lens.

\begin{figure}
\centering 
\centerline{\psfig{file=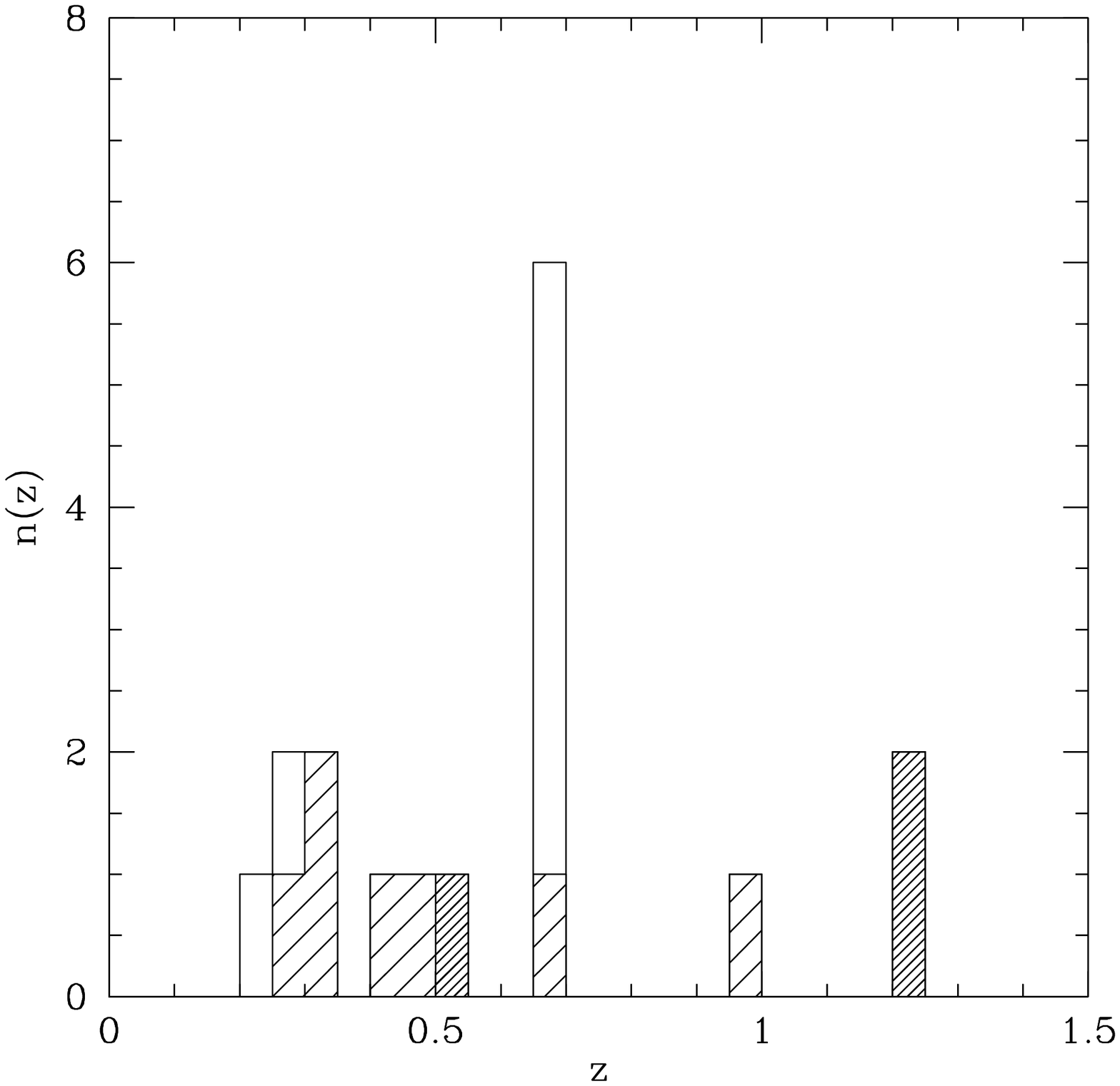,width=8.0cm}}
\caption{The redshift distribution of our KX sample.  Heavy shading denotes
the QSOs, light shading the narrow emission line galaxies and no shading
for the absorption line galaxies.  Note that most of the early type galaxies
are at $z\simeq0.66$.}
\label{nz}
\end{figure}

The remaining identifications are all galaxies, with roughly equal
numbers of early and late type spectra (by early and late we mean
spectra that are dominated by absorption and emission lines
respectively).  The early type objects are on average redder.  The
redshift distribution of the identified galaxies and QSOs is shown in
Fig. \ref{nz}.  While the narrow emission line galaxies are
distributed from $z\simeq0.15$ to $1.0$, 5 out of 7 of the early type
galaxies lie at $z\simeq0.66$, suggesting the presence of a galaxy
cluster at that redshift.

\begin{table*}
\begin{center}
\caption{$K\leq19$ galaxies observed in addition to our KX survey.
The format is the same as that in Table \ref{data_table1}.  The `?'
after the \oiii~ for KG3 indicates that the features associated with
the \oiii~ line are questionable due to the peculiar line ratios of the
two \oiii~ lines.  The `(abs)' after an \mgii~ indicates that this line
is seen in absorption.}
\tabcolsep=4pt
\begin{tabular}{lcccccccccccl}
\hline 
name & RA(J2000) & Dec.(J2000) & U & B & V & R & J & K & S/G$_{\rm K}$ & z & ID &
spectral features\\
\hline
KG1  &  3 32 10.62  & --27 47 02.0  &24.58  & 24.77  &23.43  &22.21   &20.19&   18.42  &0.07   &0.000   & ?     &   \\                             	     
KG2  &  3 32 10.81  & --27 46 28.0  &  -    &   -    & -     &24.57   &20.18&   18.34  &0.03   &0.000   & ?     &   \\				   	     
KG3  &  3 32 11.45  & --27 46 50.2  &24.39  & 24.28  &22.57  &21.86   &19.86&   18.54  &0.43   &0.577   & gal(e)   &   \oii,H,K,H$\gamma$,\oiii~?\\          
KG4  &  3 32 11.55  & --27 47 13.2  &22.84  & 23.33  &22.39  &21.56   &19.64&   17.74  &0.03   &0.000   & ?     &   \\			 		     
KG5  &  3 32 14.87  & --27 47 38.5  &25.16  & 25.46  &23.79  &22.80   &20.43&   18.90  &0.45   &0.000   & ?     &   \\			        	     
KG6  &  3 32 21.79  & --27 44 42.4  &21.82  & 22.41  &21.40  &20.96   &19.90&   18.97  &0.01   &0.295   & gal(e)   &   \oii,H,K,H$\delta$,H$\beta$,\oiii,H$\alpha$\\	     
KG7  &  3 32 22.02  & --27 46 56.2  &22.78  & 23.91  &22.64  &21.49   &19.02&   17.37  &0.04   &0.670   & gal(a)   &   H,K,H$\delta$,G,H$\gamma$\\	     
KG8  &  3 32 22.63  & --27 44 26.0  &22.06  & 22.55  &21.82  &20.94   &19.02&   17.39  &0.03   &0.738   & gal(e)   &   \mgii(abs),\oii,H,K,H$\delta$,G,H$\gamma$\\  
KG9  &  3 32 28.04  & --27 46 39.5  &22.26  & 22.18  &20.68  &19.86   &17.94&   16.23  &0.08   &0.247   & gal(a)   &   H$\beta$,Mgb,NaD,H$\alpha$,\nii \\    
KG10 &  3 32 34.16  & --27 47 12.4  &21.85  & 22.28  &21.30  &20.80   &19.74&   18.38  &0.10   &0.227   & gal(e)   &   H$\beta$,\oiii,H$\alpha$,\sii\\       
KG11 &  3 32 39.92  & --27 47 15.3  &22.43  & 23.34  &22.74  &22.21   &20.54&   18.91  &0.01   &1.097   & gal(e)   &   \mgii(abs),\oii,H,K,H$\delta$\\	     
KG12 &  3 32 42.59  & --27 43 47.4  &22.76  & 23.58  &22.25  &21.68   &21.51&   18.82  &0.00   &0.000   & ?     &   \\					     
KG13 &  3 32 42.85  & --27 46 05.9  &22.05  & 22.60  &21.89  &20.88   &19.36&   17.65  &0.04   &0.000   & ?     &   \\					     
KG14 &  3 32 45.66  & --27 45 54.9  &21.65  & 22.30  &21.18  &20.50   &19.53&   18.00  &0.01   &0.337   & gal(e)   &   \oii,H,K,H$\delta$,G,H$\gamma$,H$\beta$,\oiii,\\ 
KG15 &  3 32 45.69  & --27 44 05.9  &22.20  & 22.93  &22.15  &21.12   &19.91&   18.33  &0.02   &0.000   & ?  &   \hspace{1.3cm} NaD,H$\alpha$,\nii,\sii     \\  
KG16 &  3 32 46.60  & --27 47 08.9  &22.15  & 22.51  &21.46  &20.85   &19.98&   18.90  &0.00   &0.128   & gal(e)   &   NaD,H$\alpha$,\sii\\	     
\hline												      						          
\label{data_table3}
\end{tabular}
\end{center}
\end{table*}

\section{Discussion}\label{disc_sec}

\subsection{Blue candidates, $V-J<3$}

We can draw some preliminary conclusions on the basis of the
spectroscopy completed so far.  Only two sources brighter than
$K=19.5$ were found to be UVX and not KX.  The UV, optical and IR
colours of these two objects suggest that they are Galactic stars and
spectroscopy has confirmed this in one case.  The three confirmed QSOs
also show a UV excess.  Thus, within our sample, KX selection appears
to contain all the QSOs which would have been selected via the UVX
method, as we find no UVX QSOs which are not also KX. Of the 13 KX
candidates with $V-J<3$, 11 were  observed, and 9 have been
identified.  The effective area of this sample is then 48 arcmin$^2$
multiplied by $9/13$. (We obtain a very similar effective area if we
consider the identified fraction of objects with a stricter stellarity
cut $>0.8$).  This gives us a lower limit to the surface density of
QSOs brighter than $K=19.5$ of $325^{+316}_{-177}$ deg$^{-2}$, the
lower limit allowing for the possibility that the red sample contains
one or more QSOs.

We can calculate the number of QSOs expected in this field, based on
previous results from blue selected surveys.  Taking the result of
Boyle, Jones \& Shanks (1991), who find a density of 76.8 QSOs per
square degree at $z<2.9$ and $B<22$, an average ($B-K$)=3.0 and a
faint end number counts slope (extrapolated) of 0.3, we expect to find
$\sim1.4$ QSOs in our $48$ arcmin$^2$ field.  Given the small numbers
and the uncertainties in this calculation we conclude that our data
are consistent with this.

Five of the KX candidates are narrow emission line galaxies.  These
could be obscured AGN, where the broad line region is enshrouded in
dust. However, a more likely explanation, is that they are simply star
forming galaxies.  The current spectra are not of sufficient quality
to differentiate between these two possibilities.  However, galaxy
evolution models (e.g. see Eisenhardt et al. 2000) suggest that late
type galaxies should lie in this region of colour-colour space.
Comparison with the deep {\it Chandra} X-ray imaging data soon to be
available in this field can answer this question.

The QSO pair at $z=1.22$ is unexpected given the small size of the sample,
and the large angular separation ($>2'$) is significantly larger than any
confirmed gravitationally lensed pair.  The discovery of a galaxy
cluster at $z=0.66$ may provide a foreground lens for this source,
however a lens of $\sim2'$ separation would typically require a
cluster with a velocity dispersion $\sim1500\kms$ (assuming a simple
singular isothermal sphere potential) which is exceptionally large.
Again, the question of this unlikely event will be resolved
with the availability of the deep {\it Chandra} data, which will easily 
detect hot gas from the cluster if it is sufficiently massive to
strongly lens a background QSO.

\subsection{Red candidates, $V-J\ge3$}

Of the 41 KX candidates with $V-J\ge3$, 22 were observed, and
8 have been identified, all as galaxies. The effective area of the
search for red QSOs is therefore 9.4 arcmin$^2$. (This could be
considered an underestimate, since QSOs are easier to identify than
early-type galaxies \---\ which appear to dominate the red sample \---\ so
the 14 objects with unidentified spectra are most likely to be
galaxies.) Our spectroscopy of red candidates therefore gives only a
weak upper limit on the number of red QSOs of  1150 deg$^{-2}$
(2$\sigma$). 

To conclude, our preliminary spectroscopic results confirm that UVX
selection is contained by KX selection, and our small sample provides
a measurement of the surface density of QSOs brighter than
$K=19.5$. However our results do not yet provide useful statistics on
the number of red QSOs. We plan to obtain near-infrared spectroscopy,
combined with further optical spectroscopy, and an analysis of the
{\it Chandra} image, in pursuit of this goal. As anticipated (and as also
encountered with UVX surveys) contamination by galaxies is a problem
for faint optical/near-infrared surveys for QSOs. With forthcoming
wide-field near-IR surveys to moderate depths $K_{\rm lim}\simeq17$,
where contamination from galaxies is much less pronounced, the KX
method is an obvious choice to construct unbiased samples of QSOs.

\section*{acknowledgements}

This paper was prepared using the facilities of the STARLINK node at
the Imperial College of Science, Technology and Medicine.  We
thank the ESO Imaging Survey team for making their data publicly
available.  SMC acknowledges the support of PPARC.

{}

\end{document}